\documentclass[aps,pre,showpacs,floatfix,superscriptaddress]{revtex4}
\usepackage{amsmath}
\usepackage{amssymb}
\usepackage{graphicx}
\usepackage{bm}
\usepackage{mathrsfs}
\usepackage{subfigure}

\newcommand{\arctanh}{\text{arctanh}}

\begin{document}

\title{Ultra-short solitons and kinetic effects in nonlinear metamaterials}  

\author{Mattias Marklund} 
\affiliation{Department of Physics, Ume{\aa} University, SE--901 87 Ume{\aa},
Sweden}
\affiliation{Centre for Fundamental Physics, Rutherford Appleton Laboratory,
Chilton, Didcot, Oxfordshire, UK} 

\author{Padma K.\ Shukla}
\affiliation{Institut f\"ur Theoretische Physik IV, Fakult\"at f\"ur
  Physik und Astronomie,  Ruhr-Universit\"at Bochum, D--44780 Bochum,
  Germany} 
\affiliation{Centre for Fundamental Physics, Rutherford Appleton Laboratory,
Chilton, Didcot, Oxfordshire, UK}
  
\author{Lennart Stenflo}
\affiliation{Department of Physics, Ume{\aa} University, SE--901 87 Ume{\aa},
Sweden}

\date{January 23, 2006}

\begin{abstract}
We present a stability analysis of a modified nonlinear Schr\"odinger
equation describing the propagation of ultra-short pulses in negative refractive
index media. Moreover, using methods of quantum statistics, 
we derive a kinetic equation for the pulses, making it possible to 
analyze and describe partial coherence in metamaterials.  
It is shown that a novel short pulse
soliton, which is found analytically, can propagate in the medium. 
\end{abstract}

\maketitle

As early as in 1945, Mandelstam studied the electromagnetic properties of 
materials exhibiting a negative index of refraction \cite{Mandelstam1,Mandelstam2}.
The concepts of negative permittivity $\varepsilon$ and permeability $\mu$ and 
the consequences of negative group velocity were moreover given attention by Pafomov 
\cite{Pafomov} and Agranovich and Ginzburg \cite{Agranovich-Ginzburg}, 
and have since then been discussed by several authors, e.g.  \cite{Veselago,Agranovich1,Agranovich2,Zhang}.
Recently, such materials have been produced \cite{Smith-etal,Shelby-etal}, 
and experimental verifications of previous theoretical issues are currently 
attracting much interest (see, e.g. \cite{Pendry,Ramakrishna} for a review). 
From a sharp resonance in the material response to the applied external field,
one may obtain negative $\varepsilon$ and $\mu$.    
The normal procedure for obtaining negative-index-of-refraction materials is to put together 
two structured materials that both have negative permittivity and negative permeability, 
such that the resulting composite material has a negative refractive index 
\cite{Ramakrishna}. Since such materials may have a nonlinear response as well,
a nonlinear metamaterial can be constructed \cite{DAguanno-etal,Scalora-etal}

In the present Brief Report, we shall consider ultra-short electromagnetic 
pulses in left-handed materials \cite{Scalora-etal}.
We investigate the stability properties of a pulse, and show
that the metamaterial can influence its dynamics in significant ways. Moreover, the
effects of partial pulse coherence is studied using the Wigner function approach.
It is also shown that the dynamical equation for modulated short pulses admits a novel 
localized envelope soliton solution, which should be experimentally verified.

Given the electric field $E(z,t)\exp(ik_0z - i\omega_0t)$ of a short electromagnetic pulse, 
the evolution of the pulse envelope $E$ in the slowly varying envelope limit, 
i.e.\  $k_0 \gg (\partial_z - 2(\partial k_0/\partial\omega_0)\partial_t)$, 
is governed by (e.g.\ \cite{Scalora-etal})

\begin{equation}\label{eq:nlse}
  i\frac{\partial E}{\partial z} + \frac{\alpha}{2}\frac{\partial^2E}{\partial t^2} 
  + \kappa_1(1 - \kappa_2 I)I E - i\sigma\frac{\partial}{\partial t}(I E) = 0 ,
\end{equation}
where we have introduced the parameters $\alpha = -(\partial^2k_0/\partial\omega_0^2)$,
$\kappa_1 = \beta\mu\chi^{(3)}/2n$, $\kappa_2 = \mu\chi^{(3)}/4n^2$, and 
$\sigma = \chi^{(3)}[\mu/2V_gn^2 - (\gamma + \mu)/2n]$. Moreover, 
$\beta = 2\pi\omega_0/\omega_p$, $\omega_p$ is the plasma frequency, 
$\chi^{(3)}$ is the third order susceptibility
of the medium, $-\alpha$ is the group velocity dispersion (GVD), 
$V_g = (\partial k_0/\partial\omega_0)^{-1}$ is the group velocity,
$\mu$ is the magnetic permeability, $n$ is the refractive index, 
$\gamma =  \partial(\omega_0 n)/\partial\omega_0$, and the intensity is
given by $I = |E|^2$.

There is a stationary solution $E(t,z) = E_0\exp(i\Delta k z)$ of Eq.\
(\ref{eq:nlse}), corresponding to a nonlinear phase shift

\begin{equation}
  \Delta k = \kappa_1(1 - \kappa_2I_0)I_0 ,
\end{equation}
where $I_0 = E_0^2$.  Linearizing around this stationary solution, according to
$E(t,z) = [E_0 + E_1(t,z)]\exp(i\Delta k z)$, where $|E_1| \ll E_0$,
Eq.\ (\ref{eq:nlse}) gives

\begin{equation}
    i\frac{\partial E_1}{\partial z} + \frac{\alpha}{2}\frac{\partial^2E_1}{\partial t^2} 
  + \kappa_1(1 - \kappa_2 I_0)I_0 (E_1 + E_1^*) 
  - i\sigma I_0\frac{\partial}{\partial t}(2E_1 + E_1^*) = 0 .
  \label{eq:linear}
\end{equation}
By making the ansatz $E_1 = E_{1r} + iE_{1i}$, and assuming a harmonic dependence
in the form $E_{1r}, E_{1i} \propto \exp(iKz - i\Omega t)$, we can split Eq.\ 
(\ref{eq:linear}) into its real and imaginary parts, and obtain the dispersion relation

\begin{equation}
  K = - 2\sigma\Omega I_0 \pm \Omega\left[ 
    (\alpha\kappa_1\kappa_2 + \sigma^2)I_0^2 -  \alpha\kappa_1I_0
    + \frac{\alpha^2}{4}\Omega^2 
  \right]^{1/2}
\end{equation}  
Letting $K = -2\sigma\Omega I_0 - i\Gamma$, where $\Gamma > 0$,
we thus have the modulational instability growth rate

\begin{equation} \label{eq:modgrate}
  \Gamma = \Omega\left[  
    \alpha\kappa_1I_0 - (\alpha\kappa_1\kappa_2 + \sigma^2)I_0^2
    - \frac{\alpha^2}{4}\Omega^2
  \right]^{1/2}
\end{equation}
The effects of the material parameters $\alpha, \kappa_1, \kappa_2$,
and $\sigma$  on the modulational instability growth rate (\ref{eq:modgrate}) 
can be divided into a number of cases. We note that in the coherent case
the effect of the derivative nonlinearity is always to suppress the modulational 
instability.

First, in the case of positive GVD, i.e.\ $\alpha < 0$,  
there is no growth in a focusing medium ($\chi^{(3)} > 0$; $\kappa_1 > 0$) 
if the higher order nonlinearities through $\kappa_2$ are neglected. If the
quintic nonlinearities are taken into account for focusing media, we see from 
(\ref{eq:modgrate}) that a positive growth rate is possible, given that 
$\kappa_1\kappa_2 > 0$. Since $\kappa_1/\kappa_2 = 2\beta n$, we
will thus have growth for normal and focusing media, where both
the third order susceptibility and the refractive index are positive. 
In focusing metamaterials with positive GVD this is no longer possible,
and the modulational instability thus disappears.
If we instead have a defocusing medium, i.e.\ $\chi^{(3)} < 0$, this
implies $\kappa_1 < 0$, and $\alpha\kappa_1 > 0$. Thus, growth is 
possible even if $\kappa_2 = 0$. For normal bulk media, 
$\alpha\kappa_1\kappa_2 < 0$ which increases the modulational
instability growth rate . However, for left-handed metamaterials 
we have $\alpha\kappa_1\kappa_2 > 0$, which implies a
suppression of the modulational instability.

Secondly, there is the case of negative GVD, so that $\alpha > 0$. 
If we have a focusing medium, so that $\chi^{(3)} > 0$ and 
$\kappa_1 > 0$, the effect of a metamaterial in which $\kappa_2 < 0$,
is to increase the growth rate (\ref{eq:modgrate}), while normal media
tend to inhibit the modulational instability. However,
if the medium is defocusing, the coherent modulational instability is not
possible in a normal medium, while in left-handed metamaterials 
we can obtain a positive growth rate from (\ref{eq:modgrate}).

Thus, we see from the cases above that the 
interplay between the material coefficients is complex, and
requires careful consideration when studying the stability 
properties of waves in metamaterials. 

In order to take the effects of partial incoherence into account,
we define the space-time correlation function for the electric field as
$\widetilde{F}(z_+,z_-,t_+,t_-) = \langle E^*(z_+,t_+)E(z_-,t_-)\rangle$, 
where $z_{\pm} = z \pm \zeta/2$, $t_{\pm} = t \pm \tau/2$, and the 
angular brackets denote the ensemble average. 
The two-point correlation function for the field can
be used to analyze, for example, self-trapping of light and incoherent soliton propagation in 
optical structures and nonlinear fibers \cite{Mitchell-etal,Christodoulides-etal,Krolikowski-etal}.
Here we will use the Fourier transform of the correlation function, as this will give 
a generalized kinetic description for the quasi particles.
The Wigner distribution function of the pulse 
is given by \cite{Wigner,Mendonca}

\begin{equation}\label{eq:spacetimewigner}
  F(z,t,k,\omega) = \frac{1}{(2\pi)^2}\int\,d\zeta\,d\tau\,
  e^{i(k\zeta - \omega\tau)}\widetilde{F}(z_+,z_-,t_+,t_-) ,
\end{equation}
with the inverse $
  \widetilde{F}(z_+,z_-,t_+,t_-) = \int\,dk\,d\omega\,e^{-i(k\zeta - \omega\tau)}F(z,t,k,\omega)
$ such that

\begin{equation}\label{eq:intensity}
  I(z,t) = \int\,dk\,d\omega\, F (z,t,k,\omega) .
\end{equation}

Thus, from Eq.\ (\ref{eq:nlse}) the evolution equation for the Wigner 
function (\ref{eq:spacetimewigner}) corresponding to the envelope field $E$ is
(see also Ref.\ \cite{Besieris-Tappert})

\begin{eqnarray}
  && 
   \alpha\omega\partial_t F - \partial_z F + 
    2\kappa_1(1 - \kappa_2I) I \sin\left( \tfrac{1}{2}\stackrel{\leftarrow}{\partial_t}%
    \stackrel{\rightarrow}{\partial_{\omega}} \right) F 
\nonumber \\ && \qquad 
    + \sigma\left\{  
      \partial_t\left[ I \cos\left( \tfrac{1}{2}\stackrel{\leftarrow}{\partial_t}%
    \stackrel{\rightarrow}{\partial_{\omega}} \right) F \right] 
      - 2\omega I\sin\left( \tfrac{1}{2}\stackrel{\leftarrow}{\partial_t}%
    \stackrel{\rightarrow}{\partial_{\omega}} \right) F  \right\}
    =  0,
\label{eq:Wigner2}
\end{eqnarray}
where we have made a Wigner transformation over the time domain.
Here the arrows denote the direction of operation, and the operator
functions are defined in terms of their respective Taylor expansion. 
The system of equations (\ref{eq:intensity}) and (\ref{eq:Wigner2}) 
determines the evolution of short partially incoherent pulses in nonlinear 
meta-materials. 

In the low-frequency limit, we retain only the lowest order terms
in the $\sin$ and $\cos$ operators in Eq.\ (\ref{eq:Wigner2}),
to obtain the Vlasov-like equation

\begin{eqnarray}
%  && 
   \alpha\omega\partial_t F - \partial_z F + 
    \kappa_1\partial_t[(1 - \kappa_2I) I] \partial_{\omega} F 
%\nonumber \\ && \qquad 
    + \sigma\left[ 
      \tfrac{1}{2}\partial_t\left( I F \right) 
      - \omega(\partial_t I)\partial_{\omega}F  \right]
    =  0,
\label{eq:Vlasov}
\end{eqnarray}

In order to analyse the modulational instability and the effects of the 
terms due to nonzero $\sigma$ and $\kappa_2$, we write 
$F(z,t,\omega) = F_0(\omega) + F_1(\omega)\exp(iKz - i\Omega t) + \mathrm{c.c.}$, where 
$\mathrm{c.c.}$ denotes the complex conjugate, and $|F_1| \ll F_0$.
Expanding Eq.\ (\ref{eq:Wigner2}) in terms of this ansatz, and using Eq.\ (\ref{eq:intensity}), 
we obtain 

\begin{eqnarray}
  &&
  1 = \frac{1}{\alpha\Omega}\int_{-\infty}^{\infty}\,d\omega\,\Big\{ \left[ 
    \kappa_1(1 - \kappa_2 I_0) - \sigma(\omega + \Omega/2)\right]%
    F_0(\omega - \Omega/2) 
  \nonumber \\ && \qquad   
    - \left[ \kappa_1(1 - \kappa_2 I_0) - \sigma(\omega - \Omega/2)\right]%
    F_0(\omega + \Omega/2) \Big\}
    \left[ \omega + (K  + \sigma\Omega I_0)/\alpha\Omega \right]^{-1} ,
\label{eq:dispersion}
\end{eqnarray}
where $I_0 = \int\,d\omega\,F_0(\omega)$. Equation (\ref{eq:dispersion})
represents the NDR for a short electromagnetic pulse, where the pulse may have spectral 
broadening and partial incoherence. 

In the case of a mono-energetic pulse, we have 
$F_0(\omega) = I_0\delta(\omega - \Omega_0)$, where $\Omega_0$ 
corresponds to a frequency shift of the background plane wave solution, 
and the NDR (\ref{eq:dispersion}) gives 

\begin{equation}\label{eq:dispersion-mono}
  K = -(2\sigma I_0 + \alpha\Omega_0)\Omega \pm
    \Omega \left[  (\alpha\kappa_1\kappa_2 + \sigma^2)I_0^2 -  \alpha\kappa_1I_0
    + \frac{\alpha^2}{4}\Omega^2  + \alpha\sigma\Omega_0I_0 \right]^{1/2} .
\end{equation}

In practice however, the wave envelope will always suffer from perturbations 
due to various noise sources, e.g.\ thermal spread in the material. 
A noisy environment may cause the
pulse field to attain a random component in its phase. Thus, 
if the phase $\varphi(x)$ of the electric field varies stochastically, such that the ensemble
average of the phase satisfies \cite{Loudon}

\begin{equation}
  \langle \exp[-i\varphi(t + \tau/2)]\exp[i\varphi(t - \tau/2)]\rangle = \exp(-\Omega_T|\tau|) ,
\end{equation}
the background Wigner distribution is given by the Lorentzian spectrum
\begin{equation}
  F_0(\omega) = \frac{I_0}{\pi}\frac{\Omega_T}{(\omega - \Omega_0)^2 + \Omega_T^2},
\end{equation}
where $\Omega_T$ corresponds to the width of the spectrum. Then, the NDR (\ref{eq:dispersion}) takes the form

\begin{equation}
  1 = -I_0\Omega\frac{2\sigma\left[ K + \sigma I_0\Omega 
    + \frac{\alpha}{2}\Omega(\Omega_0 - i\Omega_T) \right] 
    + \alpha\kappa_1(1 - \kappa_2 I_0)\Omega}%
      {(K + \sigma I_0\Omega + \Omega_0 - i\Omega_T)^2 - \frac{\alpha^2}{4}\Omega^4} ,
\end{equation}
which has the solution
\begin{equation}\label{eq:sol}
  K = -\left[ 2\sigma I_0 + \alpha(\Omega_0 - i\Omega_T) \right]\Omega 
    \pm \Omega\left[  (\alpha\kappa_1\kappa_2 + \sigma^2)I_0^2 -  \alpha\kappa_1I_0
    + \frac{\alpha^2}{4}\Omega^2 
    + \alpha\sigma(\Omega_0 - i\Omega_T)I_0
  \right]^{1/2} .
\end{equation}
This solution generalizes (\ref{eq:dispersion-mono}) to the case
of a random phase background envelope field. The expression (\ref{eq:sol}) clearly 
shows that the width gives a nontrivial contribution to the NDR. 
We note that when $\sigma = 0$, we may define the growth rate $\kappa$
according to $K = -\alpha\Omega_0\Omega - i\Gamma$, and 
the width $\Omega_T$ then gives rise to a Landau like damping due to  
(\ref{eq:sol}).

\begin{figure}[ht]
  \includegraphics[width=0.8\columnwidth]{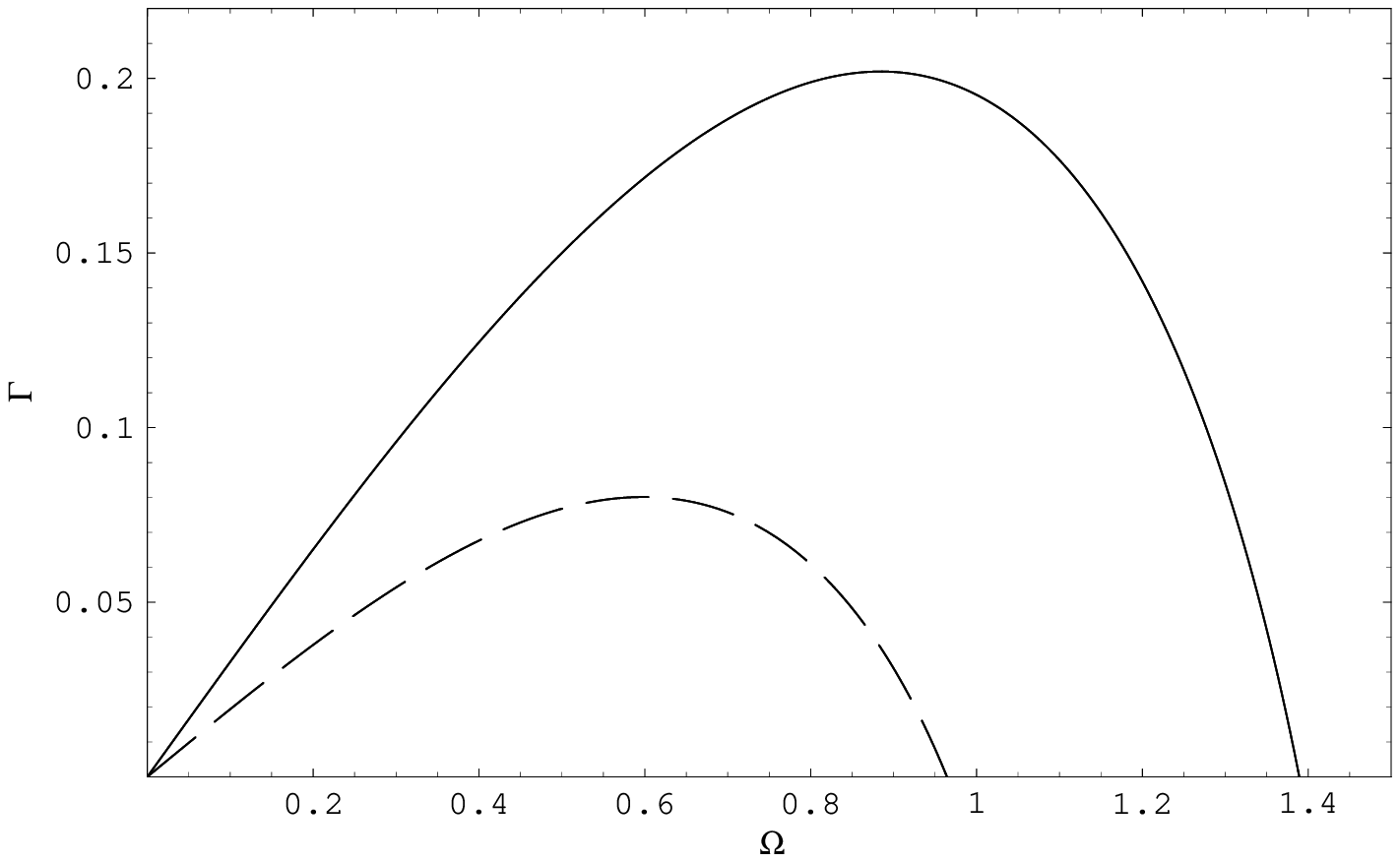}
  \caption{The modulational instability growth rate plotted
  for $I_0 = 0.5$, using $\sigma = 0.5$, $\Omega_T = 0.1$,
  and $\kappa = -0.5$ (upper curve) and $\kappa = 0.5$
  (lower curve). We have here normalized the growth rate by $\kappa_1I_0$ and the frequency by $\sqrt{|\kappa_1|I_0/\alpha}$, and correspondingly for the parameters in Eq.\ (\ref{eq:sol}), thus giving dimensionless quantities. The larger growth rate for metamaterials
  is clearly seen.}
\label{fig:2}
\end{figure}

We may normalize the dispersion relation (\ref{eq:sol}), assuming
the GVD to be negative, i.e.\ $\alpha > 0$,  according to 
$I_0 \rightarrow \kappa_1 I_0$, $\Omega_T \rightarrow \sqrt{\alpha}\,\Omega_T$,
$\Omega \rightarrow \sqrt{\alpha}\,\Omega$,  
$\sigma \rightarrow \sigma/(\kappa_1\sqrt{\alpha})$, and 
$\kappa = \kappa_2/\kappa_1 = 1/2n\beta$. The dispersion relation
(\ref{eq:sol}) then reads
\begin{equation}\label{eq:sol-norm}
  K = -\left( 2\sigma I_0 + \Omega_0 - i\Omega_T \right)\Omega 
    \pm \Omega\left[ (\kappa + \sigma^2)I_0^2 - I_0 + \tfrac{1}{4}\Omega^2 
      +\sigma(\Omega_0 - i\Omega_T)I_0 
  \right]^{1/2} .
\end{equation}
Letting $K = \mathrm{Re}\,K - i\Gamma$, where $\mathrm{Re}$ denotes
the real part, we can solve Eq.\ (\ref{eq:sol-norm}) for the growth rate 
$\Gamma$. Thus, the general perturbation exhibits oscillatory growth or damping. 
The effect of a finite $\Omega_T$ is in general to reduce the growth rate, but 
it may in certain cases broaden the spectral instability region.  We note that $\sigma$ 
and $\kappa$ have both positive and negative values, depending on whether
the material is a normal bulk medium, or if it allows for a negative index of 
refraction. For the case of a normal bulk material, we have 
$\kappa > 0$, and the higher order nonlinear correction
will thus reduce the growth rate \cite{Scalora-etal}. However,
for metamaterials we have $\kappa < 0$, altering the dynamics 
of Eq.\ (\ref{eq:nlse}) in nontrivial ways. 
Thus, for metamaterials, the modulational instability growth rate 
will differ in measurable ways from the case of normal bulk media. 
We also note that for the modulational instability growth rate, the sign of 
$\sigma$ is irrelevant. Since the full solution is symmetric
with respect to $\sigma\Omega$, requiring $\Omega > 0$ 
makes the choice of sign for $\sigma$ unimportant in this case. 
In Fig.\ \ref{fig:2} we have plotted the growth rate for both positive and 
negative values of $\kappa$, so that the difference can clearly be seen. 
Thus, it is reasonable to conclude that soliton formation and filamentation
of ultra-short pulses can be much more effective in metamaterials as compared to
normal bulk media.

\begin{figure}[ht]
  \subfigure[]{\includegraphics[width=.48\textwidth]{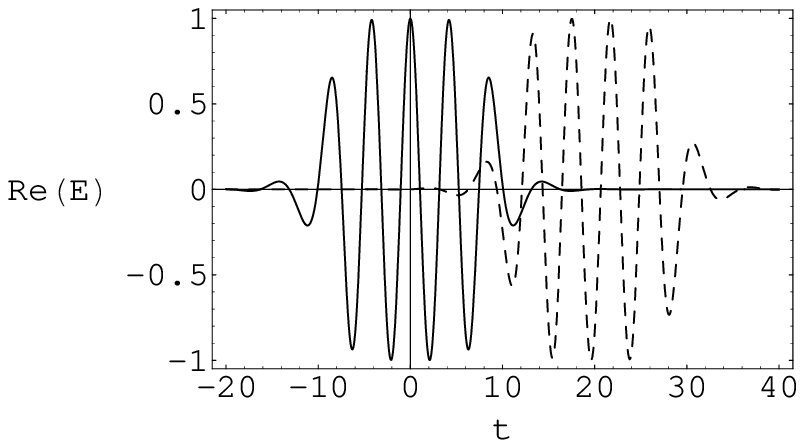}}
  \subfigure[]{\includegraphics[width=.48\textwidth]{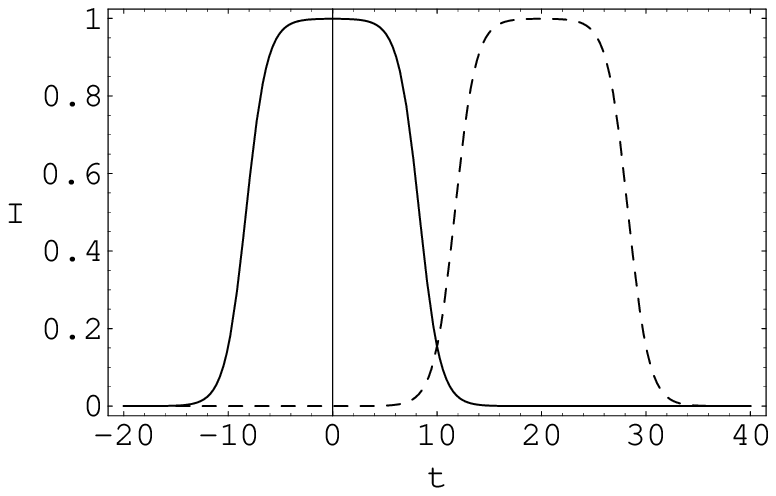}}
  \caption{The characteristic features of the soliton solution (\ref{eq:soliton}). 
  The real part of the solution (\ref{eq:soliton}) is plotted in panel (a), while the 
  intensity $I = |E|^2$ is plotted in panel (b), both for two different values of $z$. 
  We have normalized the solution
  so that $E \rightarrow \sqrt{a}E/E_0$, while time is normalized by $\sqrt{|\kappa_1|E_0^2/\alpha}$.}
  \label{fig:soliton}
\end{figure}

In fact, using the techniques of previous work (e.g.\ 
\cite{Pereira-Stenflo,Malomed1,Malomed2,Kishiba-etal,Conte-Musette,%
Afanasjev,Malomed-etal,Filho-etal}), one can obtain a novel exact analytical 
solution to Eq.\ (\ref{eq:nlse}) representing bright solitons in
the form

\begin{equation}\label{eq:soliton}
  E(t,z) = \frac{E_0  \exp[i(Kz - \Omega t + \varphi)] }{\sqrt{ a + \cosh^2[k(vt - z + z_0)]}}%
  \exp\left\{ i\delta\,\arctanh\left[\left( \frac{a}{1 + a} \right)^{1/2}\tanh[k(vt - z + z_0)] \right] 
  \right\} ,
\end{equation}
where $\varphi$ is a constant phase and $z_0$ is a spatial displacement,
and we require $a \geq 0$. Furthermore 

\begin{equation}
  \delta = \frac{3\sigma E_0^2}{2\alpha vk\sqrt{a(1 + a)}}, \quad 
  \Omega = -\frac{1}{\alpha v}, \quad
  a = \frac{E_0^2(\sigma + \alpha\kappa_1v)}{2\alpha^2k^2v^3} - \frac{1}{2}, 
  \, \text{ and } \,
  \quad
  E_0^4 =  \frac{3\alpha^4k^4v^6}{\Lambda} ,
\end{equation}
where $\Lambda = -\alpha\kappa_1v[8\kappa_2\alpha^2k^2v^3 
  - 3(\kappa_1\alpha v + 2\sigma)] + 3\sigma^2(1 + \alpha^2k^2v^4)$.
For $a \rightarrow 0$ we obtain the solitary waves found in Ref.\
\onlinecite{Radhakrishnan-etal}, if the material parameters satisfy $\sigma^2 = 
8\alpha\kappa_1\kappa_2/3$, while $a \gg 1$ corresponds to 
a flat-top wave form (see Fig.\ \ref{fig:soliton}). We also not that a nonzero 
value of $\sigma$ requires a travelling soliton solution (see Fig.\ \ref{fig:soliton}).

The width of the pulse is characterized by $1/k$. It is however limited by the slowly varying envelope approximation used in deducing the basic Eq.\ (1), and limits how short the pulse can be \cite{PRL}. This means that the parameter $k$ has to be much smaller than the pump wave number $k_0$. 

The issue of stability of solitary solutions is an interesting but complex topic 
(see e.g.\ \cite{Crasovan-etal,Malomed-etal2}).
Bright soliton solutions to the one-dimensional cubic nonlinear Schr\"odinger equation
are known to be unstable to transverse perturbations, but it is also known that competing
nonlinearities, occuring in cubic-quintic systems, can stabilize soliton solutions 
\cite{QuirogaTeixeiro-Michinel}. Our equation (\ref{eq:nlse}) contains such cubic-quintic
nonlinearites, as well as a derivative nonlinearity. It is therefore not unreasonable
to conjecture that the solution (\ref{eq:soliton}) is stable. However, in order to strictly
answer the question of stability of (\ref{eq:soliton}) much further research is needed.  

To summarize, we have examined the modulational instability and localization of 
an ultra-short electromagnetic pulse that is governed by a modified nonlinear Schr\"odinger
equation, which is also valid for negative refractive index media. 
We have considered both coherent and incoherent
pulses and derived nonlinear dispersion relations that depict the growth rates of the 
modulational instabilities. The effect of partial coherence of the electromagnetic 
pulse is to reduce the instability growth rate. Furthermore, we have found that 
modulationally unstable pulses can appear as a new type of light envelope solitons 
whose profiles are found analytically and analyzed numerically. Localized light pulses can 
propagate without distortion and should be observable in metamaterials which have negative 
index of refraction.

\ \\

\acknowledgments
This research was supported by the Swedish Research Council 
through the contract No. 621-2004-3217. 

%\newpage

%-------------------------
\newpage
%-------------------------

%------------------------
%\newpage
%------------------------

\end{document}